\documentclass[11pt,prd,aps,preprintnumbers,nofootinbib,superscriptaddress,tightenlines,onecolumn,notitlepage]{revtex4}

\usepackage{amssymb}
\usepackage{amsmath}
\usepackage[main=english]{babel}
\usepackage{microtype}
\usepackage{enumitem}
\usepackage{graphicx} 
\usepackage{physics}
\usepackage[dvipsnames]{xcolor}
\graphicspath{{figures/},{gallery/}}
\usepackage{caption}
\usepackage{tikz}
\usepackage{tikz-3dplot}
\usetikzlibrary{3d,perspective}
\usetikzlibrary{positioning}
\usetikzlibrary{calc}
\usetikzlibrary{arrows.meta}
\usetikzlibrary{shapes.geometric}
\usepackage{standalone}

\usepackage{subcaption} 
\usepackage{svg} 

\usepackage[colorlinks=true,citecolor=blue,linkcolor=blue,urlcolor=blue]{hyperref}
\usepackage{footnotebackref}

\usepackage{diagbox}
\usepackage{booktabs}

\usepackage{amsmath,amsfonts}
\usepackage{mathrsfs}
\usepackage{slashed}
\usepackage{bm}
\usepackage{bbold}

\usepackage{dcolumn}
\usepackage{bm}

\pdfoutput=1
\usepackage{lipsum}
\usepackage[normalem]{ulem}
\usepackage{braket}
\usepackage{comment}
\usepackage{bm}
\allowdisplaybreaks 

\usepackage{color,amsmath,amssymb,amsfonts,amsbsy}
\usepackage{bbm}
\usepackage{bm}
\usepackage[dvipsnames]{xcolor}
\usepackage{epsfig}
\usepackage{sidecap}
\usepackage{lipsum}
\usepackage{listings}
\usepackage{tabularx}
\usepackage{listings}
\usepackage{inconsolata}
\usepackage{dcolumn}
\usepackage{placeins} 

\usepackage{ulem}

\def\empile#1\over#2{\mathrel{\mathop{\kern 0pt#1}\limits_{#2}}}

\begin{document}

\newcommand*{\Pavia}{Dipartimento di Fisica, Universit\`a di Pavia, via Bassi 6, I-27100 Pavia, Italy}
\newcommand*{\INFN}{INFN, Sez.~di Pavia, via Bassi 6, I-27100 Pavia, Italy}
\newcommand*{\PSU}{Division of Science, Penn State University Berks, Reading, Pennsylvania 19610, USA}
\newcommand*{\Saclay}{Universit\'e Paris-Saclay - CEA - IRFU, F-91191 Gif-sur-Yvette, France}
\newcommand*{\ODU}{Department of Physics, Old Dominion University, Norfolk, VA 23529, USA }
\newcommand*{\JLab}{Jefferson Lab, Newport News, Virginia 23606, USA}

\author{Alessandro Bacchetta}
\thanks{Electronic address: alessandro.bacchetta@unipv.it -- \href{https://orcid.org/0000-0002-8824-8355}{ORCID: 0000-0002-8824-8355}}
\affiliation{\Pavia}
\affiliation{\INFN}
\author{Matteo Cerutti }
\thanks{Electronic address: matteo.cerutti@cea.fr -- \href{https://orcid.org/0000-0001-7238-5657}{ORCID: 0000-0001-7238-5657}}
\affiliation{\Saclay}
\author{Leonard Gamberg}
\thanks{Electronic address: lpg10@psu.edu -- \href{https://orcid.org/0000-0002-5112-1713}{ORCID:0000-0002-5112-1713}}
\affiliation{\PSU}
\author{Richard Whitehill}
\thanks{Electronic address: rwhit08@odu.edu -- \href{https://orcid.org/0000-0001-5007-2473}{ORCID: 0000-0001-5007-2473}}
\affiliation{\ODU}
\affiliation{\JLab}

\title{Phenomenological Estimate for the 
Longitudinal Unpolarized Structure Function\\ in Semi-inclusive Deep Inelastic Scattering}

\begin{abstract}

We present a phenomenological study of the longitudinal unpolarized
structure function $F_{UU,L}$ in semi-inclusive deep inelastic scattering
(SIDIS) and its ratio with the transverse structure function,
$R_{\, \rm SIDIS}=F_{UU,L}/F_{UU,T}$. We study their transverse-momentum
dependence, combining for the first time low- and high-transverse-momentum
contributions across the kinematic range relevant to SIDIS measurements.
At low transverse momentum, $F_{UU,L}$ is power suppressed, making it sensitive to
next-to-next-to-leading-power dynamics in the unpolarized sector.
We find that this contribution can be sizable, with
$R_{\, \rm SIDIS}\simeq 0.3$ at low transverse momentum and hard scale, producing a correction of the order of $10\%$ to transverse-momentum-integrated observables.
Although these estimates remain subject to theoretical
uncertainties, our results indicate that accounting for $F_{UU,L}$ is
likely to be important for the interpretation of SIDIS multiplicities
and motivate dedicated measurements of $R_{\, \rm SIDIS}$ to better constrain the longitudinal contribution.
\end{abstract}

\preprint{JLAB-THY-26-4985}

\maketitle

\section{Introduction}

Deep-inelastic lepton--nucleon scattering (DIS) has for decades been a primary tool for uncovering the internal structure of hadrons and providing access to 
collinear parton distribution functions (PDFs), which encode the longitudinal momentum distributions of quarks and gluons. It is well known that (neglecting parity-odd terms) the unpolarized DIS cross section is described by two structure functions, $F_{T}$ and $F_{L}$, corresponding to the exchange of transversely and longitudinally polarized virtual photons, respectively.
The longitudinal structure function can be accessed experimentally by measuring the longitudinal-to-transverse cross-section ratio $R = \sigma_L/\sigma_T$~\cite{HallC:2025}. 
This observable has been measured with moderate precision~\cite{BCDMS:1989ggw,Dasu:1988ms,Whitlow:1990gk,NewMuon:1996fwh,H1:1996zad,E143:1998nvx,Gehrmann:1999xn,H1:2010fzx}, and there is a broad agreement in the literature that $R\simeq 0.25$, a value that is also supported by phenomenological studies based on collinear PDFs~\cite{H1:2010fzx}.
From a theoretical perspective, the inclusive
longitudinal structure function $F_L$ first appears at $\mathcal{O}(\alpha_s)$.
It is therefore {\it perturbatively suppressed}, but remains a leading-power 
quantity in collinear factorization (i.e., it is not suppressed by inverse powers of the hard scale).

In more recent years, semi-inclusive deep-inelastic scattering (SIDIS), in which a hadron is detected in coincidence with the scattered lepton, has enabled access to the partonic transverse-momentum degrees of freedom within hadrons.
Beyond the inclusive Bjorken-$x$ and hard-scale ($Q^2$) dependence, the SIDIS structure functions contain additional
kinematic 
dependence on the momentum of the observed hadron $P_h$, conventionally parameterized by the hadron energy fraction $z=P\cdot P_h/P\cdot q$ together with its component $P_{h\perp}$ transverse to the photon and target momenta. Alternatively, theoretical results are often shown as functions of the component $q_T$ of the photon momentum transverse to the target and produced hadron momenta~\cite{Mulders:1995dh,Diehl:2005pc,Bacchetta:2008xw,Boussarie:2023izj}. In this work, we consider both the low- and high-transverse-momentum regions, which may be expressed in terms of $P_{h\perp}$ or, equivalently, the corresponding photon transverse momentum $q_T$.

For an unpolarized beam and an unpolarized target, the azimuthally-integrated SIDIS cross section is described by two structure functions, $F_{UU,T}$ and $F_{UU,L}$, which generalize the corresponding DIS structure functions~\cite{Mulders:1995dh,Bacchetta:2006tn}.
Most recent analyses of transverse-momentum differential SIDIS data 
have focused on $F_{UU,T}$ since it is predicted to dominate by power counting~\cite{Mulders:1995dh,Bacchetta:2006tn} in the low transverse momentum region, $q_{T} \ll Q$.
On the contrary, $F_{UU,L}$ has received little attention, and its role in transverse-momentum-dependent phenomenology remains largely unexplored. 

In the low-transverse-momentum region, $q_T \ll Q$, TMD factorization applies, and at leading power in the hard scale $Q$, the $F_{UU,T}$ structure function admits a description in terms of transverse-momentum-dependent (TMD) parton distribution and fragmentation functions \cite{Mulders:1995dh,Bacchetta:2006tn}.
In this regime, $F_{UU,L}$ is explicitly power suppressed relative to $F_{UU,T}$ and 
receives next-to-next-to-leading power (NNLP) contributions from kinematic corrections behaving as~\cite{Bacchetta:2008xw}
\begin{align}
&F_{UU,L} \;\sim\; \frac{q_T^2}{Q^2}\, F_{UU,T}\,
\end{align}
in addition to contributions from higher-twist TMD distributions~\cite{Wei:2016far}.
A complete factorization 
framework incorporating all sources of power-suppressed contributions is not yet available (see Refs.~\cite{Bacchetta:2019qkv,Ebert:2021jhy,Gamberg:2022lju,Rodini:2023plb} for works along this direction).

First phenomenological estimates of the ratio of longitudinal to transverse SIDIS structure functions based on the parton model suggest that $F_{UU,L}$ may contribute at the level of $30\%$ in certain
kinematic regions~\cite{Accardi:2023chb}.
This observation has motivated renewed theoretical interest in $F_{UU,L}$. Recent work on power corrections to SIDIS structure functions has shown that kinematic contributions to $F_{UU,L}$ can be generated from twist-two TMD distributions while retaining the essential structure of TMD evolution~\cite{Piloneta:2025jjb}. Complementary studies based on rapidity factorization have derived NNLP contributions to the unpolarized SIDIS structure functions, including contributions involving both unpolarized and Boer--Mulders/Collins TMDs~\cite{Balitsky:2026nux,Balitsky:2017flc}.

In the large transverse momentum region where $q_T \sim Q$, the observed transverse momentum is generated perturbatively through hard partonic radiation~\cite{Koike:2006fn,Bacchetta:2008xw}. 
The appropriate theoretical description is then provided by collinear factorization, in which the SIDIS cross section is computed as a fixed-order expansion in the strong coupling $\alpha_s$, 
known up to $\mathcal{O}(\alpha_s^2)$. Thus, the small- and large-$q_T$ regions are described by theoretically distinct mechanisms: intrinsic and perturbatively evolved transverse-momentum dependence at small $q_T$  and hard QCD radiation at large $q_T$.

A quantitatively reliable description across the full transverse-momentum range requires a consistent connection between these two regimes. Such a description is provided, for $F_{UU,T}$, by matching the 
TMD contribution at small $q_T$ to the fixed-order collinear result at large $q_T$, for example through the CSS $W+Y$ construction~\cite{Collins:1984kg,Sun:2013hua,Boglione:2014oea,Collins:2016hqq}. For $F_{UU,L}$, however, the situation is not  established. The twist-4 TMD terms at small $q_T$ should in principle match onto twist-2 collinear terms at large $q_T$ for $F_{UU,L}$, but problems with applying the standard approach were first highlighted in Ref.~\cite{Boer:2006eq} in the context of Drell-Yan scattering and reemphasized in Ref.~\cite{Bacchetta:2008xw} in the context of SIDIS. 
In this work, we do not attempt to solve these problems but propose a reasonable way to provide smooth phenomenological estimates of $F_{UU,L}$ across the full $q_T$ spectrum.

The relevance of $F_{UU,L}$ extends beyond the structure function itself. Transverse-momentum-dependent SIDIS measurements are commonly presented in terms of hadron multiplicities, namely ratios of semi-inclusive to inclusive DIS cross sections. 
In global TMD analyses, the inclusive DIS denominator contains both the
transverse and longitudinal structure functions, $F_T$ and $F_L$,
while the contribution of $F_{UU,L}$ to the numerator is usually neglected. 
A consistent treatment of this term is therefore relevant for the interpretation of hadron multiplicities and, in particular, for extractions of unpolarized TMD parton distribution and fragmentation functions~\cite{Gonzalez-Hernandez:2018ipj,Wang:2019bvb,Boglione:2016bph,Collins:2016hqq,Sun:2014dqm,Boglione:2014oea,Piacenza:2020sst,Gonzalez-Hernandez:2023iso,Bacchetta:2022awv}. The same issue is relevant for angular modulations and spin asymmetries, whose experimental extraction involves ratios to the unpolarized SIDIS cross section.

Despite its phenomenological importance, experimental information on the transverse-momentum dependence of $F_{UU,L}$ remains very limited. In SIDIS, the only available measurements are integrated in transverse momentum and are affected by large experimental errors~\cite{Bebek:1975pn,Bebek:1976wv,Bebek:1977pf}.  New dedicated measurements are being pursued at Jefferson Lab and planned for future facilities~\cite{Accardi:2023chb}. Such measurements would provide direct information on the transverse-momentum dependence of the longitudinal contribution and offer an important test of the theoretical descriptions discussed above.

In this study, we develop a 
description of $F_{UU,L}$ that connects the low-$q_T$ TMD region to the high-$q_T$ collinear regime. 
In the absence of a complete TMD factorization and resummation formalism for $F_{UU,L}$, we adopt a phenomenological strategy.
We develop models for the low-$q_T$ behavior of $F_{UU,L}$ and investigate their connection to the fixed-order collinear result at large transverse momentum. The resulting matching should be regarded as exploratory, providing insight into the potential impact of longitudinal-photon contributions on SIDIS multiplicities and on the ratio $R_{\, \rm SIDIS}$, while simultaneously highlighting the need for a rigorous factorization and resummation treatment of $F_{UU,L}$.

This paper is organized as follows.
In Sec.~\ref{s:theory} we review the theoretical framework for describing the transverse-momentum dependence of SIDIS structure functions, including both TMD and collinear factorization. In Sec.~\ref{s:pheno} we present our phenomenological estimates for $F_{UU,L}$ and $R_{\, \rm SIDIS}$, discussing the matching. 
Finally, in Sec.~\ref{s:conclusions} we summarize our findings and discuss their implications for future experimental and theoretical studies.

\section{Theoretical Framework}
\label{s:theory}

In this section, we describe the theoretical framework used to calculate the transverse and longitudinal structure functions in SIDIS and to connect their descriptions in the small- and large-transverse-momentum regions.
The theoretical separation of these regions is organized in terms of the photon transverse momentum $q_T$\footnote{Results in the SIDIS literature are often presented in terms of the hadron transverse momentum $P_{h\perp}$. We discuss the definitions of these two transverse momenta and their relation in App.~\ref{a:kinematics}.}.
This transverse momentum can span a broad range, from the nonperturbative region $M \sim q_T \ll Q$,
where $M$ denotes an intrinsic nonperturbative scale where TMD dynamics are relevant, to the perturbative region
$M \ll q_T \sim Q$, where hard QCD radiation generates the transverse momentum and collinear factorization applies.

The low-transverse momentum limit ($q_T \ll Q$) of the structure functions (often referred to as ``$W$ term'') can be calculated using TMD factorization theorems \cite{Collins:1981uw,Ji:2004wu,Ji:2004xq,Collins:2011qcdbook,Boussarie:2023izj}.
In this work, we focus on the unpolarized, azimuthal-integrated transverse and longitudinal structure functions $F_{UU,T}$ and $F_{UU,L}$.
At leading power, only the transverse structure function is non-zero, and is expressed as a convolution 
of the unpolarized TMD PDFs and FFs $f_1$ and $D_1$ as
\begin{align}
\label{eq:FUUT_LP_TMD}
    F_{UU,T} &(x, z, q_T, Q) 
    = \mathcal{C}[ \omega f_1 D_1 ] 
    = \sum_{a} x e_a^2 \int \dd[2]{\vb*{p}_{T}} \dd[2]{\vb*{k}_{T}} \, f_1(x,\vb*{p}_T^2) D_1(z,\vb*{k}_T^2) \delta^{(2)}( \vb*{p}_{T} - \vb*{k}_{T} + \vb*{q}_{T} ) \,
,\end{align}
where the sum is over all quark flavors $a$, 
$\bm{p}_{T}$ and $\bm{k}_{T}$ 
are the transverse momenta of the parton before and after the photon interaction, respectively (for more details, see Ref.~\cite{Bacchetta:2006tn}).
For brevity,  we suppress the dependence on the renormalization group and rapidity scales $\mu$ and $\zeta$, respectively,  of the TMDs.
Recently, subleading power corrections to these structure functions have been studied~\cite{Piloneta:2025jjb,Balitsky:2026nux}
.
While the details differ, they find NNLP corrections to the transverse structure function and a NNLP expression for the longitudinal structure function.
In particular, they obtain a contribution to $F_{UU,L}$ of the form $\mathcal{C}[(4 p_T^2 / Q^2) f_1 D_1]$, as predicted also in Refs.~\cite{Bacchetta:2008xw}, and additionally, a Boer--Mulders contribution is found~\cite{Balitsky:2026nux,Piloneta:2025jjb}.

In the high transverse momentum region ($q_T \gg M$), the structure functions are computed within collinear factorization at fixed order in perturbation theory (referred to as the ``$FO$ term''):
\begin{align}
\label{eq:stf-coll-factorization}
    F_{UU,T/L} &= \frac{1}{Q^2} \frac{\alpha_s}{2 \pi^2 z} \int_{\xi_{\rm min}}^{1} \frac{\dd{\xi}}{\xi} \int_{\zeta_{\rm min}}^{1} \frac{\dd{\zeta}}{\zeta^2} \, \delta\Big( \frac{q_T^2}{Q^2} - \frac{( 1 - \hat{x})(1 - \hat{z})}{\hat{x} \hat{z}} \Big) \nonumber \\
    &\hspace{1cm} \times \sum_{ij} C^{(ij)}_{UU,T/L}(\hat{x},Q^2,\hat{z},q_T) f_{i/N}(\xi,Q^2) D_{h/j}(\zeta,Q^2)
,\end{align}
where the sum over indices $i,j$ is over all parton species, $\hat{x} = x/\xi$ and $\hat{z} = z/\zeta$ with $\xi$ and $\zeta$ being the collinear momentum fractions of the incoming and fragmenting partons, respectively.
Importantly, it is known that the $FO$ contribution from $F_{UU,L}$, like $F_{UU,T}$, is leading power~\cite{Mendez:1978zx,Bacchetta:2008xw}.
In this work, we use the coefficient functions at leading order~\cite{Mendez:1978zx,Bacchetta:2008xw} so that
\begin{subequations}
\label{eq:FO-coeff-funcs}
\begin{align}
C^{(qq)}_{UU,T}
&= 2 C_F e_q^2 \Bigg\{ 2\hat{x}\hat{z}
 + \Bigg[
    2 + \frac{(1-\hat{x})^2 + (1-\hat{z})^2}
             {\hat{x}\hat{z}}
   \Bigg]
   \frac{Q^2}{q_T^2}
   \Bigg\}
\quad &
C^{(qq)}_{UU,L}
&= 8 C_F e_q^2 \hat{x}\hat{z}
\label{eq:Cqq}
\\[1ex]
C^{(qg)}_{UU,T}
&= 2 C_F e_q^2 (1-\hat{z})
   \Bigg\{
   2\hat{x}
   + \frac{1+(\hat{x}-\hat{z})^2}
          {\hat{x}\hat{z}^2}
     \frac{Q^2}{q_T^2}
   \Bigg\}
\quad &
C^{(qg)}_{UU,L}
&= 8 C_F e_q^2 \hat{x}(1-\hat{z})
\label{eq:Cqg}
\\[1ex]
C^{(gq)}_{UU,T}
&= 2 T_R e_q^2 (1-\hat{x})
   \Bigg\{
   4\hat{x}
   + \frac{
      1 - 2\hat{x}(1-\hat{x})
        - 2\hat{z}(1-\hat{z})
     }
     {\hat{x}\hat{z}^2}
     \frac{Q^2}{q_T^2}
   \Bigg\}
\quad &
C^{(gq)}_{UU,L}
&= 16 T_R e_q^2 \hat{x}(1-\hat{x})
\label{eq:Cgq}
.\end{align}
\end{subequations}

In the intermediate region ($M \ll q_T \ll Q$), two situations can arise. If the TMD expression expanded in this region has the same power behavior as the fixed-order expression, the two should match~\cite{Bacchetta:2008xw}. If their power behaviors differ, they describe different physical mechanisms, and both should be taken into account. For $F_{UU,T}$, it is well known that the first situation applies, and the two expressions indeed match. For $F_{UU,L}$, the low-transverse-momentum kinematic contributions discussed above have the same power behavior as the fixed-order result. However, these contributions alone cannot fully reproduce the fixed-order result~\cite{Bacchetta:2008xw}. It remains an open question whether a complete NNLP TMD expression can be found that yields the correct matching.
Furthermore, although the $W$ and $FO$ expressions for $F_{UU,T}$ match in the intermediate-$q_T$ region, achieving a numerically well-behaved implementation of $F_{UU,T}$ has been a challenge~\cite{Boglione:2014oea,Collins:2016hqq,Gonzalez-Hernandez:2023iso,Maxia:2025zee}.

To circumvent these issues, we adopt a transition function $w(q_T)$ to interpolate between the $W$ and $FO$ terms within the intermediate region so that
\begin{align}
\label{eq:matched-stf}
    F = w(q_T) F_{W} + [1 - w(q_T)] F_{FO} \, .
\end{align}
We reserve a discussion of the exact form of the transition function for the subsequent section. 
We note that alternatives to this interpolation procedure have been discussed in the literature (see, e.g., \cite{Collins:2016hqq,Echevarria:2018qyi}).

\section{Phenomenological Estimates}
\label{s:pheno}

In this section, we present a phenomenological estimate of the longitudinal structure function $F_{UU,L}$ in SIDIS, focusing on the ratio $R_{\, \rm SIDIS}= F_{UU,L}/F_{UU,T}$, which provides a direct
probe of subleading-power TMD 
dynamics in the unpolarized sector. 
We stress that this is the first time a systematic phenomenological study of the transverse-momentum dependence of the longitudinal structure function in SIDIS has been performed, and for this reason, we take particular care in the identification of a suitable model and in the determination of the uncertainties associated with the estimate of
$R_{\, \rm SIDIS}$.
The final result will be obtained by combining the different sources of uncertainty, which will be discussed in detail in the following sections. 
The main purpose of this work is to provide a meaningful estimate of the longitudinal structure function in SIDIS, which can be used as a benchmark for future studies and to guide experimental efforts.

Our determination of $R_{\, \rm SIDIS}$ is based on a consistent use of TMD PDFs and TMD FFs grids from the MAP22 extraction~\cite{Bacchetta:2022awv}, which are used to compute the $W$-term contribution to both $F_{UU,T}$ and $F_{UU,L}$.
As described above, we use the fixed-order calculation at  $\mathcal{O}(\alpha_s)$ for both structure functions, and we use the \textsc{MMHT2014}~\cite{Harland-Lang:2014zoa} and \textsc{DSS14}~\cite{deFlorian:2014xna} sets at NLO for collinear PDFs and FFs, respectively.
To achieve a consistent matching between the $W$ and $FO$ terms we use TMD grids at NNLL accuracy. 
Although $\mathcal{O}(\alpha_s^2)$ fixed-order calculations and N$^3$LL TMD grids are available, their implementation is not straightforward and goes beyond the scope of this work.

This section is organized as follows.
In Sec.~\ref{ss:parameterizations} we discuss the different parameterizations of the small-$q_T$ behavior of the longitudinal structure function that we consider in our analysis. 
We then present a matching prescription to connect the small-$q_T$ TMD description with the high-$q_T$ collinear regime in Sec.~\ref{ss:matching_prescription}, which is crucial for obtaining a consistent estimate of $R_{\, \rm SIDIS}$ point-by-point in $q_T$.
In Sec.~\ref{ss:final_result} we identify different sources of uncertainty in our computation of 
$R_{\, \rm SIDIS}$
and estimate their size, which we use to
obtain a result for the central value and total uncertainty for  $R_{\, \rm SIDIS}$.  
The impact of our results on the $q_T$-integrated $z$-dependent SIDIS structure function $F_L$ is discussed in Secs.~\ref{ss:integral} 
as well as their impact on the description of transverse-momentum-dependent SIDIS data at kinematics accessible at fixed target facilities in Sec.~\ref{ss:impact_on_data}.
Finally, in Sec.~\ref{ss:Comparison_alternative} we compare our results with alternative approaches in the literature and assess the role of NNLP TMD contributions to $F_{UU,T}$ in the description
of $R_{\, \rm SIDIS}$.

\subsection{Tested parameterizations}
\label{ss:parameterizations}

As discussed above, the structure function $F_{UU,L}$ at low $q_T$ is expected to be suppressed by 
$q_T^2/Q^2$ (NNLP) relative to $F_{UU,T}$, and to receive contributions from kinematical and dynamical power corrections~\cite{Bacchetta:2008xw,Balitsky:2026nux,Piloneta:2025jjb}. 
In this work, we base our estimate of $F_{UU,L}$ on the model proposed in Ref.~\cite{Bacchetta:2008xw} and confirmed in Refs.~\cite{Balitsky:2026nux,Piloneta:2025jjb}, which incorporates kinematical power corrections
consistent with the expected power counting:
\begin{equation}
\label{eq:FUUL_TMD}
F_{UU,L} (x, Q^2, z, q_T) = \mathcal{C}\bigg[ \omega(p_T^2) f_1(p_T^2) D_1(k_T^2) \bigg] \, ,
\end{equation}
where
$f_1$ and $D_1$ are the unpolarized TMD PDF and FF respectively, and
 $\omega(p_T^2)$  is a weight function
chosen to ensure the correct power counting and to provide a reasonable description of the expected behavior of $F_{UU,L}$ at low $q_T$.
Note that we do not include the contribution related to Boer--Mulders and Collins functions in this work since its effect is small relative to the kinematic contribution, as observed in Ref.~\cite{Balitsky:2026nux}. 
Furthermore, no phenomenological extractions are currently available for Boer--Mulders and Collins functions within a framework 
compatible with the unpolarized TMD PDFs and FFs used in this work, particularly in terms of their evolution.

The weight function provided by Refs.~\cite{Bacchetta:2008xw,Balitsky:2026nux,Piloneta:2025jjb} for $F_{UU,L}$ is given by the quadratic form
\begin{equation}
\omega(p_T^2) = \frac{4M^2}{Q^2} \frac{p^2_T}{M^2}
\label{e:weight_std}
\end{equation}
where $M$ is the mass of the target hadron.
This parameterization generates the expected $q_T^2/Q^2$ suppression and provides a simple functional form for the weight function.
However, the effect of the weight function on the $q_T$-dependence of $F_{UU,L}$ is quite strong. 
First, the quadratic weight significantly enhances the contribution to $F_{UU,L}$ from the tails of the TMDs, which increases the magnitude of the structure function substantially.
Furthermore, it generates a considerably larger width for the $q_T$ distribution of $F_{UU,L}$ compared to $F_{UU,T}$.
These features can spoil the matching with the high-$q_T$ region, where $F_{UU,L}$ is expected to have a similar magnitude and width as $F_{UU,T}$.
For this reason, we also consider the following parameterization for the weight function:
\begin{equation}
\omega(p_T^2) = \frac{4 p^2_T / Q^2}{1 + 4 p_T^2 / Q^2} \, .
\label{e:weight_dip2}
\end{equation}
Note that this parameterization still incorporates the expected $q_T^2/Q^2$ suppression, evident by the similar small-$p_T$
behavior of Eq.~(\ref{e:weight_dip2}) and Eq.~(\ref{e:weight_std}), but at the same time it moderates the size of $F_{UU,L}$ at small $q_T$ and avoids the strong broadening of $F_{UU,L}$ into intermediate $q_T$, allowing for smoother numerical matching between the $W$ and $FO$ terms.
For this reason, in our phenomenological studies below, we use Eq.~(\ref{e:weight_dip2}).

\subsection{Matching prescription}
\label{ss:matching_prescription}

Because of the issues discussed above in forming a strict $W+Y$ construction of the unpolarized structure functions, in this analysis, we introduce a phenomenological matching prescription to connect the small-$q_T$ TMD description of the structure functions with the high-$q_T$ collinear regime, available from fixed-order calculations.
Specifically, we calculate the structure function in Eq.~(\ref{eq:matched-stf}) by introducing a transition function $w(q_T)$ so that the full structure function reproduces the $W$ term at small $q_T$ and the $FO$ at large $q_T$, while ensuring a smooth interpolation between the two at intermediate transverse momentum.
In this work, our choice for the transition function is
\begin{equation}
    w(q_T) = 
    \begin{cases}
    1, & q_T < q_T^{(1)}, \\[6pt]
    \displaystyle
    \Bigg[ 1 + c\,\exp( \alpha \frac{2s-1}{s(1-s)} ) \Bigg]^{-1},
    & q_T^{(1)} \leq q_T \leq q_T^{(2)}, \\[10pt]
    0, & q_T > q_{T}^{(2)},
    \end{cases}
    \label{e:transition_func}
\end{equation}
with
\begin{align}
\label{eq:matching-params}
\begin{aligned}
q_T^{(1)} &= 0.1\, Q+b_1 \, , 
&\qquad s &= \frac{q_T-q_T^{(1)}}{q_T^{(2)}-q_T^{(1)}} \, ,  \\[4pt]
q_T^{(2)} &= 0.25\,Q+b_2 \, , 
&\qquad c &= \exp\left(-\alpha\,\frac{2s_0-1}{s_0(1-s_0)}\right) \, .
\end{aligned}
\end{align}
The parameters $b_1$ and $b_2$ control the offsets of the transition points $q_T^{(1)}$ and $q_T^{(2)}$, respectively, between the small-$q_T$ and intermediate-$q_T$ and intermediate-$q_T$ and large-$q_T$ regions. We take $b_1 \in [0.4,0.6]$ and $b_2 \in [1.5,2.0]$ with central values at $b_1 = 0.5$ and $b_2 = 1.5$.
The additional parameters $\alpha$ and $s_0$ control the relative overlap of the \textit{W} and \textit{FO} terms and the position of equal mixing, respectively, in the intermediate region $q_T^{(1)} < q_T < q_T^{(2)}$.
For the phenomenological estimates presented below, we adopt $\alpha=1$ and $s_0=0.3$ as our central values.
We will discuss the impact of this choice on our final results in Sec.~\ref{ss:final_result}.

The choice of the transition function is arbitrary to a certain degree, and different methodologies may be considered~\cite{Collins:2016hqq,Echevarria:2018qyi,Bizon:2018foh}.
An important feature of the transition, however,  that should be preserved from the standard $W+Y$ construction is that, as the hard scale $Q$ increases, the upper bound of the small-$q_T$ region should remain near $M$ while the lower bound of the large-$q_T$ region may grow more rapidly.
We note that such a separation of regions is difficult to achieve with one-scale transition functions, such as that considered in Ref.~\cite{Liu:2021jfp}, and may cause the $W$ or $FO$ terms to extend too far out of their regions of validity.

\begin{figure}[t]
    \centering
    \includegraphics[width=\linewidth]{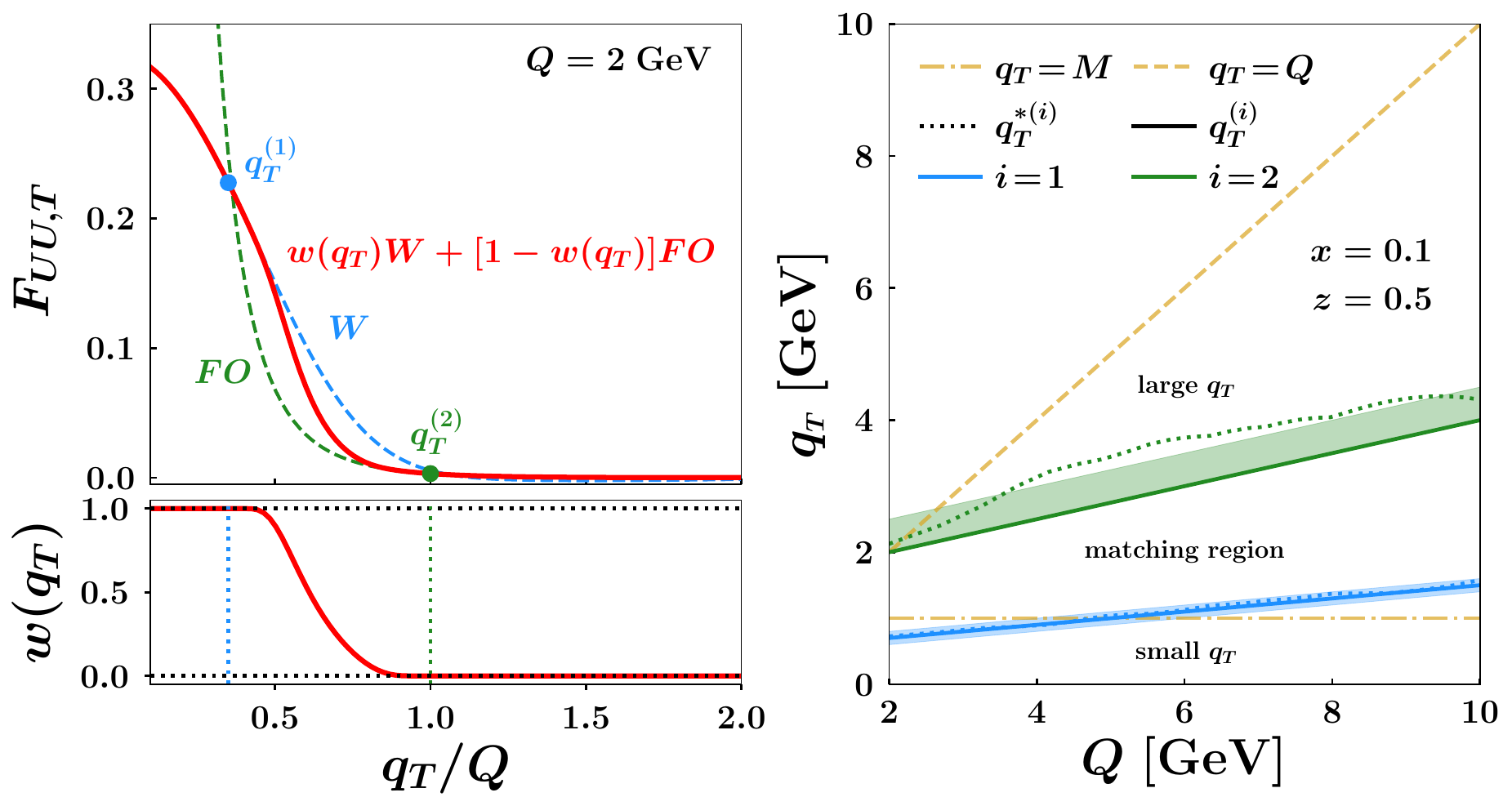}
    \caption{
    Illustration of the matching procedure used to combine the $W$ and $FO$ terms. 
    \textit{Left panel}: The $W$ (blue dashed), $FO$ (green dashed), and matched (solid red) contributions as functions of $q_T/Q$, together are shown in the upper panel with the corresponding weight function $w(q_T)$ in the lower panel. The transition points $q_T^{(1)}$ and $q_T^{(2)}$ are indicated by blue and green markers, respectively. 
    \textit{Right panel}: The intersection points $q_T^{*(1,2)}$, defined by $W(q_T)=FO(q_T)$, compared with the transition points $q_T^{(1,2)}$ as functions of $Q$ are shown alongside the reference scales $q_T=M$ and $q_T=Q$. The bands indicate the variations associated with $b_1$ and $b_2$ in Eq.~(\ref{eq:matching-params}).}
    \label{fig:matching-demo}
\end{figure}

Our matching scheme is illustrated explicitly in the left panel of Fig.~\ref{fig:matching-demo} at $x=0.1$, $z=0.5$, and $Q=2~{\rm GeV}$.
In the upper part the left panel, we show the $W$ and $FO$ terms separately as dashed green and blue lines, respectively, and we display the full structure function in red.
We also highlight the transition points of these two curves $q_T^{(1,2)}$, between which the interpolation occurs.
At the chosen kinematics $q_T^{(1)} \approx 0.4 Q \approx 0.8~{\rm GeV}$, which lies within the nonperturbative region, while the second intersection $q_T^{(2)} \approx 1.1 Q$ and lies near the canonical threshold where $FO$ calculations are expected to effectively become the full description of the cross section.
The specific form of the weight function at these kinematics is shown in the lower part of the left panel, providing a smooth interpolation between the $W$ and $FO$ terms in the region between the two intersection points.

We also show the behavior of the transition points $q_T^{(1,2)}$ defined in Eq.~\eqref{eq:matching-params} as a function of $Q$ in the right panel of Fig.~\ref{fig:matching-demo} for the same $x$ and $z$.
In addition, we include reference lines of $q_T = M$ and $q_T = Q$, which might be considered as standard qualitative boundaries from power counting for the strictly nonperturbative and perturbative regions.
It is observed that both curves for the intersections points follow a roughly linear trajectory as $Q$ increases, and the line for $q_T^{(1)}$ starts below $M$ at small $Q$ and slowly rises above this threshold at the largest $Q$ shown of $10~{\rm GeV}$.
However, this excess is well below $2M$, indicating that our $W$ term stays contained within a nonperturbative window.
On the other hand, the line for $q_T^{(2)}$ starts slightly above $Q$ and increases relatively faster than the $q_T^{(1)}$ line, although not as fast as the expected $q_T \simeq Q$.

For reference, we also show the intersection points $q_T^{*(1)}$ and $q_T^{*(2)}$, defined by $W(q_T)=FO(q_T)$, as dotted lines in the right panel of Fig.~\ref{fig:matching-demo}. We find that the former lies well within the band of the transition region defined by $q_T^{(1)}$, while the latter is generally larger than $q_T^{(2)}$, although still compatible. This supports our choice of the transition regions.

It should be noted that at arbitrary kinematics an intersection between $W$ and $FO$ terms is not guaranteed, whether this be for numeric or theoretical reasons.
In such cases, we extrapolate the intersection points from nearby kinematics where an intersection exists.
Furthermore, general caution should be taken when $Q$ becomes comparable with nonperturbative scales since the matching region collapses.

\subsection{Uncertainty estimates and final result}
\label{ss:final_result}

A crucial aspect of our analysis is the assessment of the uncertainties associated with our predictions for 
$R_{\, \rm SIDIS}$.  We identify different sources of uncertainty, which can be broadly categorized into three main groups: (i) matching uncertainty associated with the choice of the transition function $w(q_T)$; (ii) theoretical uncertainty associated with the statistical precision of the TMD and collinear PDF and FF sets; and (iii) theoretical uncertainty from the truncation of perturbative expansions at fixed order, manifesting through renormalization scale dependence of structure functions. 
For the uncertainty associated with the matching procedure, we evaluate this by varying the parameters $\alpha$ and $s_0$ in Eq.~(\ref{eq:matching-params}) within reasonable ranges around the baseline values at the central TMD and collinear PDF and FF distributions which define $F_W$ and $F_{FO}$ in Eq.~(\ref{eq:matched-stf}).
The model uncertainty is then defined as the resulting spread in the $F_{UU,L}$ and $R_{\, \rm SIDIS}$.

Statistical uncertainties from the TMD replicas and the collinear PDF and FF sets are treated separately, according to their respective contributions to the matched cross section. The TMD replicas build the $W$ term and therefore determine the associated uncertainty predominantly in the low-$q_T$ region. Similarly, the collinear PDF and FF sets are used in the calculation of the $FO$-term, and their uncertainties are therefore propagated at large $q_T$. Although the TMDs are constructed from the collinear distributions through the OPE, the dependence on the collinear PDFs and FFs at low $q_T$ is already incorporated in the spread of the TMD replicas. Consequently, we treat the statistical uncertainties associated with the TMD replicas and collinear sets as independent sources.
We then use standard error propagation to write
\begin{align}
\label{eq:F-stat-err}
    \delta F = \sqrt{ w^2(q_T) \delta F_{W}^2 + \big[ 1 - w(q_T) \big]^2 \Big[ \delta F_{FO,{\rm PDF}}^2 + \delta F_{FO,{\rm FF}}^2 \Big] }
,\end{align}
where
\begin{subequations}
\begin{align}
\label{eq:MC-err}
    \delta F_{W}^2 &= \Delta_{68\%}( \{ F_{W}^{(1)}, \hdots, F_{W}^{(N_{\rm rep})} \} ) \\
\label{eq:hess-err}
    \delta F_{FO}^2 &= \frac{1}{4} \sum_{i=1}^{(N_{\rm mem} - 1)/2} ( F_{FO}^{(i+1)} - F_{FO}^{(i)} )^2
.\end{align}
\end{subequations}
Eq.~(\ref{eq:MC-err}) simply represents the $68\%$ confidence interval for the TMD replicas, 
and Eq.~(\ref{eq:hess-err}) is the standard error propagation for hessian sets~\cite{Pumplin:2001ct}.
Note that we use Eq.~(\ref{eq:hess-err}) for the collinear PDF and FF errors separately since they are statistically independent objects.

Finally, we study the impact of missing higher-order contributions by performing scale variations of both the $W$ term and $FO$ calculation. 
For this we vary the renormalization and rapidity scales by a factor of two around their central value $\mu=\sqrt{\zeta}=Q$.
We then estimate the uncertainty from missing higher-order corrections as the envelope of these scale variations.
Note that we do not consider independent factorization scales for the $FO$ calculation since they may introduce unnecessary complications from the renormalization scheme transformation, which can impact the implementation of sum rules and renormalization group evolution as well as the matching between the $W$ and $FO$ terms~\cite{Rogers:2026qlt}.

The total uncertainty on $F_{UU,L}$ and is then defined as the sum in quadrature over all the three sources discussed above.
While this procedure may result in a conservative estimate of the uncertainty, we consider it appropriate for a first phenomenological study of this NNLP structure function.
The statistical uncertainty of $R_{\rm SIDIS}$ can be obtained from Eq.~(\ref{eq:F-stat-err}) using standard error propagation techniques for the ratio of $F_{UU,L}$ and $F_{UU,T}$.

\begin{figure}[t]
\centering
\includegraphics[width=\textwidth]{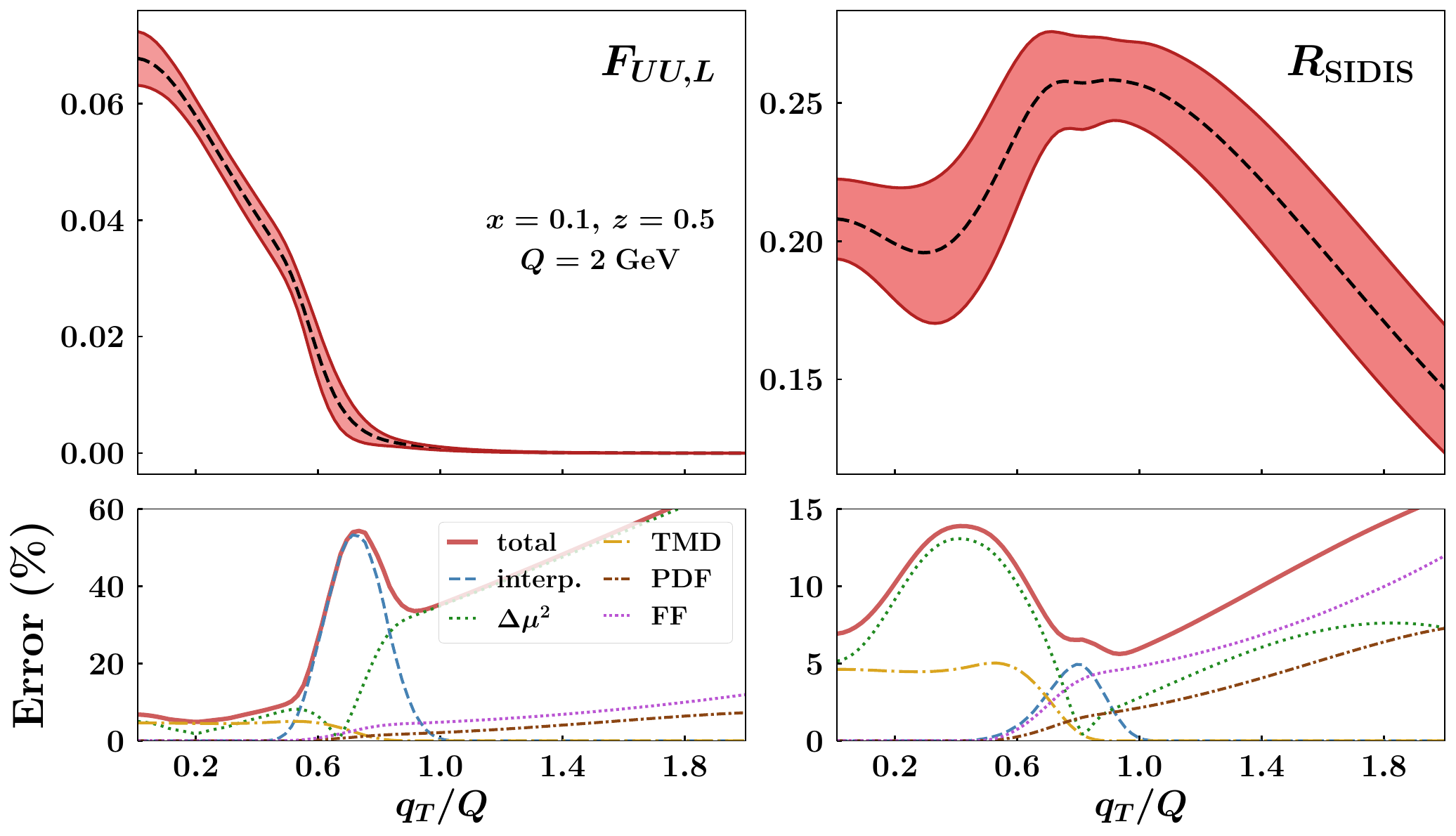}
\caption{Transverse-momentum dependence of $F_{UU,L}$ (left panels) and the ratio $R_{\rm SIDIS}$ (right panels) at $x=0.1$, $z=0.5$, and $Q^2=4$ GeV$^2$. Uncertainty bands in the upper panel are obtained by summing the errors from various sources displayed in the lower panels in quadrature.}
\label{f:uncertainties}
\end{figure}

In Fig.~\ref{f:uncertainties}, we plot $F_{UU,L}$ and $R_{\rm SIDIS}$ at $x=0.1$, $z = 0.5$, and $Q = 2~{\rm GeV}$, showing the central values and total uncertainty band in the top panels.
The total uncertainty depicted in the band is decomposed into its relative components in the bottom panels.
We observe that the matching uncertainty is quite large and provides the dominant error source for $F_{UU,L}$ at intermediate $q_T$, where the transition function is designed to interpolate between the small-$q_T$ and large-$q_T$ descriptions of the structure functions.
However, at this same intermediate region, it appears that such uncertainty cancels to a large extent in the ratio $R_{\rm SIDIS}$ and becomes comparable to the other sources.

As expected by construction, for both $F_{UU,L}$ and $R_{\rm SIDIS}$, the TMD replica uncertainty is confined primarily to the small transverse momentum region and falls off quickly as $q_T$ approaches the $FO$ region.
On the other hand, collinear hessian uncertainties start rising in the intermediate region and become sizable at the largest $q_T$.
It is therefore important to improve the precision of TMD and collinear extractions in order to reduce their impact on the total uncertainty on $F_{UU,L}$ and $R_{\rm SIDIS}$.

Missing higher-order corrections provides a sizable uncertainty at small transverse momentum and can be larger than even the TMD replica uncertainty in $R_{\rm SIDIS}$.
It also provides the dominant source of uncertainty at large transverse momentum in $F_{UU,L}$, although this appears to cancel in the ratio and becomes subleading compared to the collinear hessian uncertainties.
To mitigate the sizable effect of scale dependence, it is crucial to extend the perturbative and logarithmic accuracy of our calculations to achieve a more precise description of $R_{\rm SIDIS}$, both at low and high transverse momentum. We leave this as a future extension of this work.

\subsection{Integral over transverse momentum}
\label{ss:integral}

\begin{figure}[t]
\centering
\includegraphics[width=\textwidth]{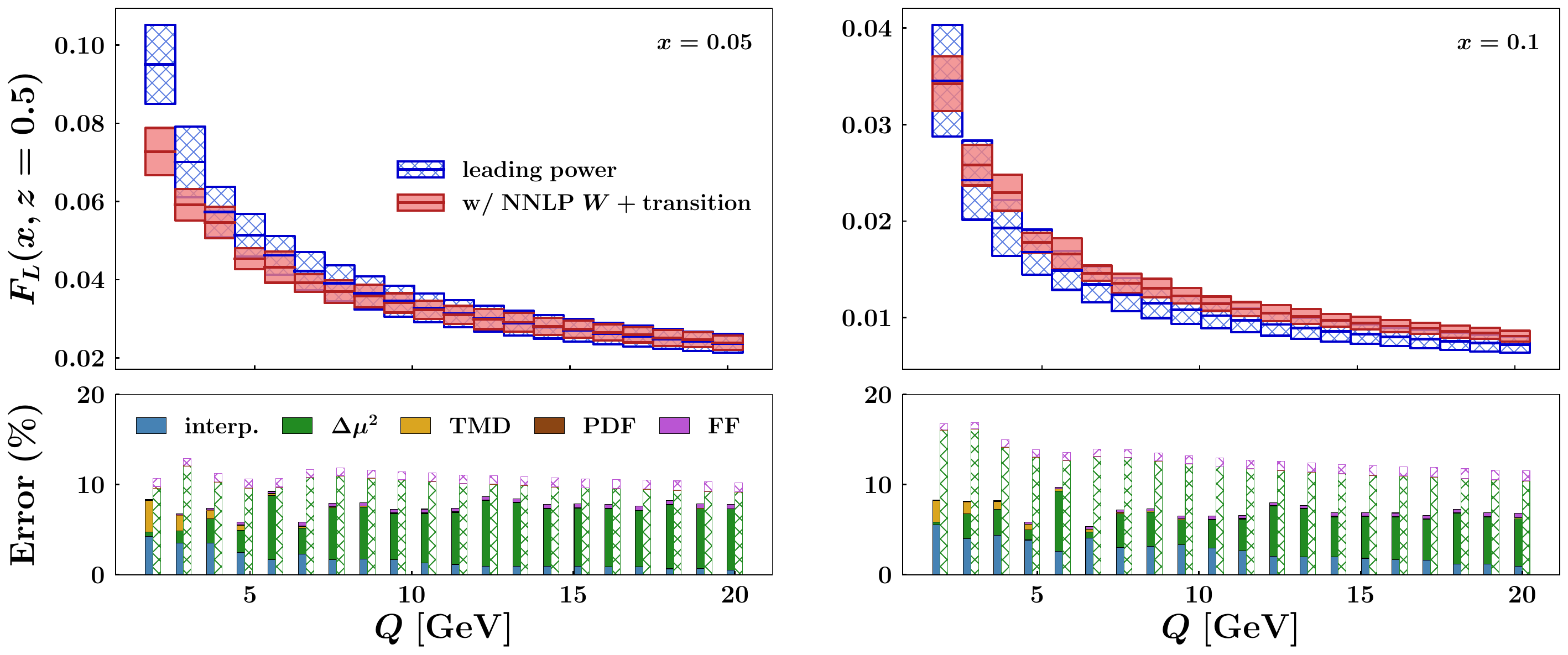}
\caption{The integrated longitudinal structure function over transverse momentum as a function of $Q$ at $x=0.1$ and $z=0.5$ (left panels) and at $x=0.01$ and $z=0.5$ (right panels). The red band corresponds to the integrated structure function obtained from our model, while the blue band corresponds to the integral of only the fixed-order calculation. In the upper panels we show the integrated structure function, while in the lower panels we show the breakdown and the size of the different sources of uncertainties.}
\label{f:integral}
\end{figure}

We now investigate the impact of our small-$q_T$ model for $F_{UU,L}$ on the semi-inclusive structure function $F_L$, obtained from $F_{UU,L}$ by integrating over the transverse momentum of the final measured hadron.
As we discussed in Sec.~\ref{ss:parameterizations}, the contribution of NNLP contributions to $F_{UU,L}$ to the integrated structure function is expected to be small, suppressed by $1/Q^2$, but can still have an impact on the description of data, especially at low $Q^2$. 
To assess this impact, the upper panel of Fig.~\ref{f:integral} compares the structure function $F_L(x,z,Q)$ obtained by integrating our matched $F_{UU,L}$ over $q_T$ (red band) with the corresponding result obtained by integrating the $FO$ contribution alone (blue band), as a function of $Q$. We have verified that the latter agrees with the numerical result of the structure function in collinear factorization.
We show the results at $x=0.05$ and $x=0.1$ in the left and right panels, respectively, both at $z=0.5$.
In both panels, the upper plot shows the integrated structure function, while the lower plot shows the breakdown and the size of the different sources of uncertainties.
We observe that the contribution of $F_{UU,L}$ to the integrated structure function is small, but it can still be sizable, especially at low $Q^2$. 
In particular, at $Q \sim 2$ GeV, the difference between the integral of our model and the fixed-order calculation can reach values on the order of $10\%$ at $x=0.05$ and $5\%$ at $x=0.1$.
This is a consequence of the expected $q_T^2/Q^2$ suppression of $F_{UU,L}$ at low transverse momentum, which is more significant at higher $Q$.
Indeed, from Fig.~\ref{f:integral} we observe that the difference between the two integrals decreases as $Q$ increases and becomes negligible at $Q \sim 10$ GeV.
We also note that the uncertainty on the integrated structure function is quite large, particularly at low $Q^2$, and that it is larger for the integral of the $FO$ calculation than for our model.
This is a consequence of the fact that the integral of the $FO$ is sensitive to the behavior of $F_{UU,L}$ at low $q_T$, where the theoretical uncertainties are larger, while our model provides a more stable result in this region.
The dominant source of uncertainty for the $FO$ calculation is given by scale variations, which generates an error on the order of $10\%$ at $x<0.05$, while it becomes larger at $x=0.1$.
Collinear FF errors, on the order of a few percent, provide another significant source of theoretical uncertainty.
For our model, instead, the relative size of different sources of uncertainties is more variable as a function of $Q$. 
At $Q=2$ GeV, the dominant sources of error are TMD replicas and interpolation uncertainty. 
Beyond this value and up to $Q \sim 10$ GeV, the uncertainty from missing higher-order corrections becomes the dominant source.
Importantly, at larger $Q$, the distribution of uncertainties for both the model and $FO$ integral as well as the size of the total uncertainty become compatible, indicating that our matching scheme is well behaved in this limit.

\subsection{Impact on description of SIDIS data}
\label{ss:impact_on_data}

\begin{figure}[h!]
\centering
\includegraphics[width=\textwidth]{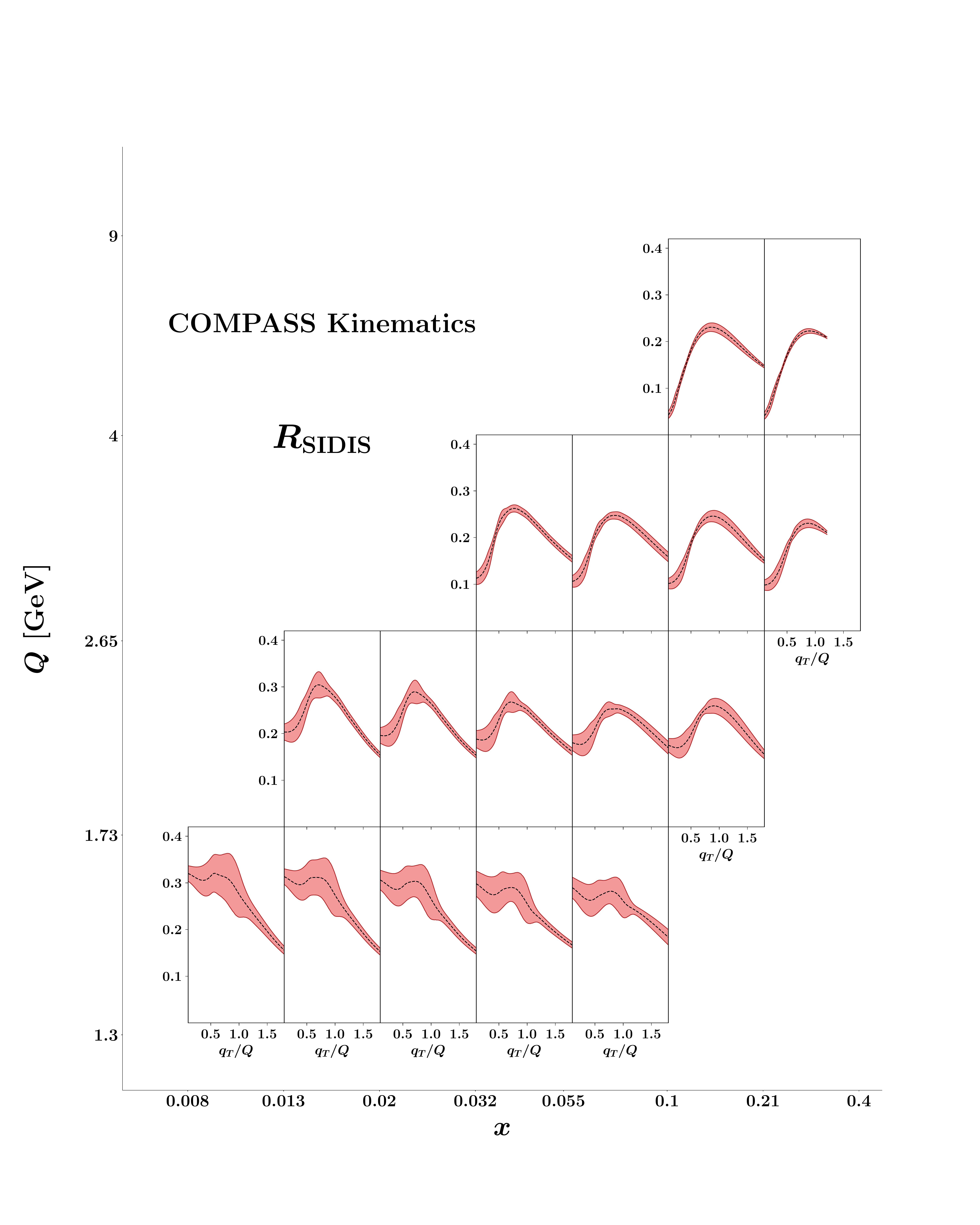}
\caption{
Theoretical prediction for $R_{\rm SIDIS}$ in the kinematic region covered by COMPASS data for the production of $\pi^+$ off a proton target. For each panel we show $R_{\rm SIDIS}$ as a function of $q_T/Q$ at $z=0.5$. Uncertainty bands are obtained following Sec.~\ref{ss:final_result}.}
\label{f:R_compass}
\end{figure}

It is important to assess the impact of our result on the description of SIDIS data included in global analyses. In particular, we focus on hadron multiplicities, which are defined as the ratio between the semi-inclusive and inclusive DIS cross sections.
Both the DIS and SIDIS cross sections are given by the sum of the transverse and longitudinal structure functions. 

In the case of integrated multiplicities in the transverse momentum of the final measured hadron, the contribution of $F_{L}$ is already included, but only the leading power calculation in collinear factorization is considered, consistent with the integral of only the $FO$ term of $F_{UU,L}$.  As we discussed in Sec.~\ref{ss:integral}, the contribution of $F_{UU,L}$ to the integrated multiplicities is expected to be small, but it can still have an impact on the description of data, especially at low $Q^2$. 
In fact, it may help to reduce the tensions between HERMES and COMPASS data in collinear FF extractions~\cite{deFlorian:2014xna,deFlorian:2017lwf,Borsa:2022vvp}.

In the case of multiplicities differential in the transverse momentum of the final measured hadron, 
as stated earlier,  
the  contribution of $F_{UU,L}$ to the numerator is  neglected in  current phenomenological studies, under the assumption that it is a correction of NNLP and therefore negligible, potentially generating a mismatch between numerator and denominator. Given the expression for the transverse momentum dependent multiplicty,  
\begin{align}
M^h(x,z,Q^2,P_{h\perp}^2)
&=
\frac{F_{UU,T}+\epsilon F_{UU,L}}
     {F_T+\epsilon F_L}
=
\frac{F_{UU,T}}
     {F_T+\epsilon F_L}
\left(1+\epsilon R_{\, \rm SIDIS}\right),
\label{eq:sidis-multiplicity}
\end{align}
where  $\epsilon$ denotes the ratio of longitudinal to transverse virtual-photon fluxes~\cite{Bacchetta:2007sz},
its inclusion may help to reduce the normalization mismatch observed in the description of the hadron multiplicities at high theoretical accuracy~\cite{Bacchetta:2022awv}, and provide more robust results in studies of angular modulations and spin asymmetries, where the denominator of the measured observables is represented by the unpolarized cross section.
Therefore, dedicated studies of $F_{UU,L}$ and  $R_{\, \rm SIDIS}$  will allow a more consistent treatment of both the numerator and denominator of hadron multiplicities, and will provide a more robust description of SIDIS data.

Here we focus here on the differential multiplicities in transverse momentum, and we study the size of $R_{\rm SIDIS}$ in the kinematic region covered by COMPASS data. 
In Fig.~\ref{f:R_compass}, we show our result for the production of $\pi^+$ off a proton target. 
For each panel, which corresponds to a specific $(x,Q)$ bin in the COMPASS kinematics, we report $R_{\rm SIDIS}$ as a function of $q_T/Q$ at $z=0.5$. 
Uncertainty bands are obtained following Sec.~\ref{ss:final_result}.
Overall, we observe that $R_{\rm SIDIS}$ is non-negligible in the kinematic region covered by COMPASS data, as it can reach values on the order of $30\%$ at low $q_T$ and low $Q$. 
Its $q_T$ dependence is strong at larger $Q$, while it flattens at low $Q$, suggesting that in the latter case it may induce an enhancement of the multiplicity normalization.
As expected, the size of $R_{\rm SIDIS}$ at low transverse momentum strongly depends on $Q$, with values on the order of $5\%$ at the highest $Q$ bin and on the order of $30\%$ at the lowest $Q$ bin. 
This is a consequence of the expected $q_T^2/Q^2$ suppression of $F_{UU,L}$ at low transverse momentum.
We also note a sizable dependence of $R_{\rm SIDIS}$ on the value of $x$, with larger values at smaller $x$, particularly for the fixed-order term. 
This may be a consequence of both the enlarged partonic kinematic phase space and the enhancement of the gluon contribution to $F_{UU,L}$ at smaller $x$.
No significant dependence on $z$ is observed, reflecting the similar $z$ dependence of $F_{UU,L}$ and $F_{UU,T}$, which largely cancels in the ratio.
We also observe that the uncertainty on $R_{\rm SIDIS}$ is quite large, particularly at low $Q$, where the model uncertainty is dominant. 
This highlights the importance of improving our knowledge of the $F_{UU,L}$ structure function, and to reduce the associated theoretical uncertainties.
In general, this suggests that the contribution of $F_{UU,L}$ plays a role in the description of SIDIS data, and it would be important to devise a strategy to include it in future phenomenological studies.

\subsection{Comparison with alternative approaches}
\label{ss:Comparison_alternative}

It is now instructive to compare our results with previous calculations available in the literature.
In particular, we focus on the comparison with the results of Refs.~\cite{Piloneta:2025jjb,Balitsky:2026nux}. The former, hereafter denoted as ``Madrid'', provides a comprehensive study of kinematic power corrections to the SIDIS structure functions in a resummation approach, and the latter (``BP'') presents a study of the same structure functions in the rapidity factorization formalism.

\begin{figure}[t]
    \centering
    \includegraphics[width=\linewidth]{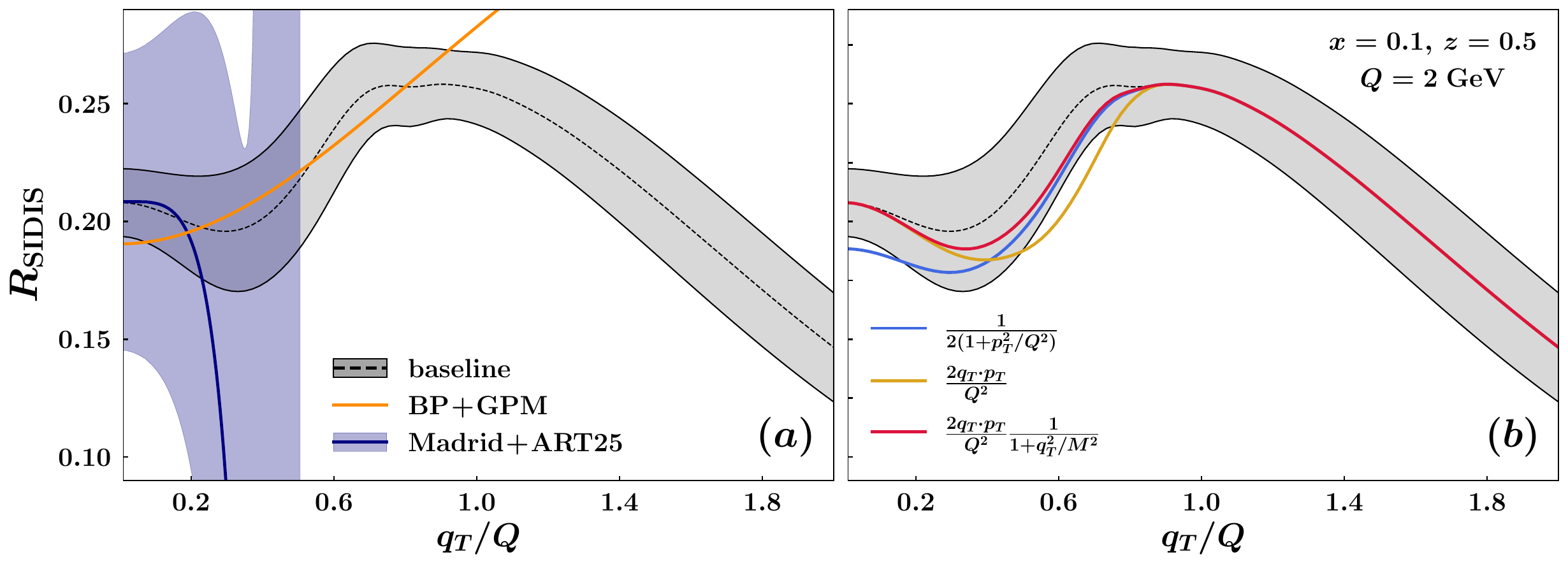}
    \caption{Theoretical predictions of $R_{\rm SIDIS}$ at $x=0.1$, $z=0.5$, and $Q=2$ GeV. \textit{Left panel} comparison of our result (gray band) with the ones of Refs.~\cite{Piloneta:2025jjb,Balitsky:2026nux} (blue band and orange curve, respectively). \textit{Right panel}: comparison of our baseline result (gray band) and the results obtained in our framework but including the Madrid, BP and our phenomenological expressions for the NNLP correction to $F_{UU,T}$ (blue, yellow, and red curves, respectively).
    }
    \label{fig:R_literature_comparison}
\end{figure}

In the left panel of Fig.~\ref{fig:R_literature_comparison}, we show a comparison between our result for $R_{\rm SIDIS}$ (gray band) and the results of Refs.~\cite{Piloneta:2025jjb,Balitsky:2026nux} (blue band and orange curve, respectively) at $x=0.1$, $z=0.5$, and $Q=2$ GeV. Note that the matching with fixed-order calculations is not considered in the Madrid and BP studies and, therefore, their results should be considered valid only at small $q_T$. 
Overall, we observe a good qualitative agreement between our result and both the Madrid and BP studies at small transverse momentum. This is particularly interesting since the three studies are based on different theoretical frameworks and use different parameterizations for the small-$q_T$ behavior of $F_{UU,L}$. This agreement suggests that the three different approaches are mutually consistent and provide a similar description of $R_{\rm SIDIS}$ at low transverse momentum.

Finally, we investigate the impact of the NNLP correction to $F_{UU,T}$ on our results. The corresponding expression obtained by the Madrid study is given by
\begin{equation}
F_{UU,T}^{\rm Madrid} = F_{UU,T}^{\rm LP} + \frac{1}{2} F_{UU,L} \, ,
\end{equation}
whereas the BP result reads
\begin{equation}
F_{UU,T}^{\rm BP} \simeq F_{UU,T}^{\rm LP} + \mathcal{C}\bigg[\frac{2\vb*{q}_T \cdot \vb*{p}_T}{Q^2} f_1 D_1\bigg] \, ,
\end{equation}
where we neglected the contribution of the Boer--Mulders TMD PDF and Collins TMD FF, which is expected to be small.
To assess the impact of this correction on our results, we compute $R_{\rm SIDIS}$ using the Madrid and BP expressions for $F_{UU,T}$ and the same TMD PDFs and FFs as in our baseline calculation. We emphasize that both expressions have different $q_T$ dependence from the LP result at intermediate transverse momentum and, are therefore are incompatible with the power counting given by the FO calculation. For this reason, we introduce an additional model for the $q_T$ dependence of the NNLP correction to $F_{UU,T}$, based on phenomenological considerations. In particular, we assume that it has a similar expression as in BP, but we multiply it by a weight function, similar to the one used for $F_{UU,L}$, to suppress its contribution at intermediate $q_T$ and to allow for a smooth matching with the FO calculation. The expression of this model is given by
\begin{equation}
F_{UU,T}^{\rm Pheno} \simeq F_{UU,T}^{\rm LP} + \mathcal{C}\bigg[\frac{2\vb*{q}_T \cdot \vb*{p}_T}{Q^2}\frac{1}{1+q_T^2/M^2} f_1 D_1\bigg] \, .
\end{equation}

In the right panel of Fig.~\ref{fig:R_literature_comparison}, we show a comparison between our baseline result for $R_{\rm SIDIS}$ (gray band) and the results obtained by using the Madrid and BP expressions for $F_{UU,T}$ (blue and green curves, respectively) at $x=0.1$, $z=0.5$, and $Q=2$ GeV. We also show the result obtained by using our model for the NNLP correction to $F_{UU,T}$ (orange curve). Overall, we observe that the impact of the NNLP correction to $F_{UU,T}$ on our results is small and does not significantly change the size of $R_{\rm SIDIS}$. In particular, we observe that the Madrid and BP expressions for $F_{UU,T}$ lead to a slightly smaller value of $R_{\rm SIDIS}$ at $q_T/Q \simeq 0.6$. We note that the BP expression for the NNLP correction to $F_{UU,T}$ vanishes at small $q_T$, while the Madrid expression is non-zero and suppresses the size of $R_{\rm SIDIS}$. Our model for the NNLP correction to $F_{UU,T}$, instead, yields a result that is in between the Madrid and BP expressions, and it is closer to our baseline result. This is expected, since the phenomenological suppression factor preserves the low-$q_T$ behaviour while damping the correction in the matching region.
In general, all these results are compatible within the uncertainty of our baseline calculation. This indicates that the inclusion of the NNLP correction to $F_{UU,T}$ does not significantly change our conclusions.

\section{Conclusions}
\label{s:conclusions}

In this work, we have presented phenomenological estimates of the longitudinal structure function $F_{UU,L}$ in unpolarized SIDIS and of the ratio $R_{\rm SIDIS}=F_{UU,L}/F_{UU,T}$, focusing on their transverse-momentum dependence.
We have combined a model for the low-$q_T$ contribution, where $F_{UU,L}$ is suppressed by $q_T^2/Q^2$ relative
to $F_{UU,T}$, with the fixed-order collinear result, where both $F_{UU,L}$ and $F_{UU,T}$ contribute at leading power.
In the absence of a complete NNLP TMD factorization framework, we have based our model on the expected power counting involving unpolarized TMD PDFs and FFs from previous extractions, and we have adopted a phenomenological interpolation between the two regions. This procedure provides a smooth description across the transverse-momentum spectrum, however it lacks a complete theoretical description of $F_{UU,L}$ and does not resolve the theoretical problems associated with the matching~\cite{Bacchetta:2008xw}. For this reason, we have also estimated the uncertainties associated with different theoretical sources, such as the interpolation, the input distributions, and missing higher-order corrections.  

Our results provide, for the first time, an estimate of the transverse-momentum dependence of $R_{\rm SIDIS}$ across the full $q_T$ spectrum, allowing us to study how its behavior changes between the low- and high-transverse-momentum regions.
In the COMPASS kinematics considered here, the ratio exhibits different transverse-momentum dependencies depending on the hard scale. At larger $Q$, it starts from small values at low $q_T$, rises rapidly, and subsequently decreases following the fixed-order calculation. At low $Q$, it starts from values on the order of $30\%$ and decreases with increasing $q_T$, approaching the fixed-order description.
The uncertainty on the ratio is sizable, particularly at low $Q$, where the input TMD and the missing-higher-order uncertainties are dominant.

After integration over transverse momentum, the difference between our estimate of the longitudinal structure function and the fixed-order result reaches approximately $5$--$10\%$ at $Q \simeq 2$ GeV in the kinematics studied and becomes negligible at larger $Q$.

These findings highlight the importance of including the contribution of $F_{UU,L}$ in the interpretation of measured hadron multiplicities differential in transverse momentum, as it can affect their predicted shape and normalization and consequently the determination of unpolarized TMD PDFs and FFs.
Such effects may be particularly relevant at the relatively low values of $Q$ encountered in fixed-target experiments, precisely where our estimates yield the largest values of $R_{\rm SIDIS}$.

We have compared our results at small transverse momentum with alternative approaches available in the literature and we have found that our results are in qualitative agreement. We have also investigated the impact of NNLP corrections to $F_{UU,T}$ on our results, finding that they have a limited effect on $R_{\rm SIDIS}$.

Overall, our results suggest that NNLP contributions in SIDIS can be sizable, although their magnitude and transverse-momentum dependence remain subject to substantial theoretical uncertainties. Dedicated measurements of $R_{\rm SIDIS}$ differential in transverse momentum are therefore needed to establish the size of this contribution and constrain its behavior. On the theoretical side, higher-order calculations and progress toward a complete NNLP factorization and resummation framework are needed to improve the reliability of the estimates and establish a systematic connection between the low- and high-$q_T$ regimes.

\begin{acknowledgments}
We would like to thank Sara  Pilo\~neta \'Alvarez,  Harut Avakian,  Alexey Prokudin,  and Alexey Vladimirov for correspondence, discussions and comments on this work.  This work was supported by
the U.S. Department of Energy (DOE)  under contract  No.~DE-SC0026320 (L.G.). L.G. also thanks  Jefferson Lab Theory Division, for their hospitality and support through the DOE Visiting Faculty Program (VFP) during which part of this work was carried out. The work of M.C. has been supported by l’Agence Nationale de la Recherche (ANR), project ANR-24-CE31-7061-01.
This material is based upon work supported by the U.S. Department of Energy, Office of Science, Office of Workforce Development for Teachers and Scientists, Office of Science Graduate Student Research (SCGSR) program (R.~M.~W.). 
The SCGSR program is administered by the Oak Ridge Institute for Science and Education (ORISE) for the DOE. 
ORISE is managed by ORAU under contract number DESC0014664. 
All opinions expressed in this paper are the author’s and do not necessarily reflect the policies and views of DOE, ORAU, or ORISE.
This material is based upon work supported by the U.S. Department of Energy, Office of Science, Office of Nuclear Physics under Contract No. 89243126CSC000213.
\end{acknowledgments}

\appendix

\section{Kinematic definitions and relating photon and hadron transverse momenta}
\label{a:kinematics}

In Sec.~\ref{s:theory}, we present the theoretical formalism for TMD and collinear factorization in terms of the photon transverse momentum $q_T$, defined to be the component of $q$ orthogonal to the $P$ and $P_h$.
Formulas and results in the SIDIS literature, though, are often 
presented in terms of the hadron transverse momentum $P_{h \perp}$, i.e., the component of $P_h$ orthogonal to $q$ and $P$, 
which is more naturally connected to experimental measurements.\footnote{Often in the literature the notation $P_{h T}$ is used instead of 
$P_{h \perp}$: we prefer to use the second notation here to make clear that $P_{h \perp}^{\mu}$ and $q_T^{\mu}$ are orthogonal to two different pairs of vectors.} The two four-vectors vectors are defined as (see also App.~A of Ref.~\cite{Piloneta:2025jjb})
\begin{align}
    q_{T}^{\mu} &= q^{\mu} - \frac{\cos{\theta_{T\perp}}}{\eta z} P_{h}^{\mu} + \frac{2x}{\gamma^2} \bigg( \frac{\cos{\theta_{T\perp}}}{\eta} - 1 \bigg) P^{\mu}, 
        \label{e:qTDef}
        \\
    P_{h\perp}^{\mu} &= P_{h}^{\mu} - \eta z \cos{\theta_{T\perp}}\, q^{\mu} + \frac{2 x z}{\gamma^2} \big( \eta \cos{\theta_{T\perp}} - 1 \big) P^{\mu},
    \label{e:PperpDef}
\end{align}
where\footnote{Note that the Lorentz invariant $\theta_{T\perp}$ represents the angle between the $T$ and $\perp$ transverse subspaces.}
\begin{align*}
\begin{alignedat}{3}
Q^2  &={}-q^2,
&\qquad\quad   x  &={}\frac{Q^2}{2P\cdot q},
&\qquad\quad z  &={}\frac{P\cdot P_h}{P\cdot q},
\\
\gamma  &={}\frac{2Mx}{Q},
&\qquad\quad  \gamma_h &={}\frac{M_h}{Qz},
&\qquad\quad  \rho   &={}\frac{q_T}{Q},
\\
    \cos{\theta_{T\perp}} &= \sqrt{ \frac{1 + \gamma^2(1 - \rho^2)}{1 + \gamma^2} }, 
    &
    \eta &= \sqrt{ \frac{1 - \gamma^2 \gamma_h^2}{1 + \gamma^2} }
.
\end{alignedat}
\end{align*}
For convenience, we use the notation
\begin{align*}
q_T &= \sqrt{-q_T^{\mu} q_{T \mu}} =\sqrt{\bm{q}_T^2},
\qquad\qquad
P_{h \perp} = \sqrt{-P_{h \perp}^{\mu} P_{h \perp \mu}} 
=\sqrt{\bm{P}_{h \perp}^2}
,\end{align*}
where $\bm{q}_T$ and $\bm{P}_{h\perp}$ represent the spatial transverse vectors of the virtual photon and observed hadron in the hadron-hadron and Breit frames, respectively.
Note that in the massless limit ($\gamma \to 0$),
the coefficients of $P^\mu$ in Eqs. \eqref{e:qTDef} and \eqref{e:PperpDef} approach $x(1-\rho^2)$ and $-xz(1+\rho^2)$, respectively.
The relation between $P_{h \perp}^{\mu}$ and $q_T^{\mu}$ can be written as
\begin{align}
    P_{h\perp}^{\mu} = - \eta z \cos{\theta_{T\perp}} \, q_T^{\mu} + \sin^{2}{\theta_{T\perp}} \, \Big( P_{h}^{\mu} - \frac{2 x z}{\gamma^{2}} P^{\mu} \Big).    
\end{align}
In the low-$q_T$ region ($\rho \to 0$), it reduces to Eq.~(2.25) of Ref.~\cite{Piloneta:2025jjb}. 
In the massless limit ($\gamma \to 0$) it reduces to Eq.~(4) in Ref.~\cite{Bacchetta:2019qkv}.
With the above expressions, it is easy to verify that 
\begin{equation}
P_{h\perp} = \eta z\,q_T \approx z \, q_T,
\end{equation}
where the last approximation is valid in the massless limit ($\gamma \to 0$).

\clearpage
\newpage
\bibliography{refs}
\end{document}